\documentstyle[12pt]{article}
\def\ve{{\bf e}}
\def\vm{{\bf m}}
\def\vk{{\bf k}}
\def\hk{\hat{k}}
\def\hm{\hat{m}}
\def\hn{\hat{n}}
\def\he{\hat{e}}
\def\hl{\hat{l}}

\def\vhm{{\bf \hm}}
\def\vhe{{\bf \he}}
\def\vhk{{\bf \hk}}

\def\L{\Lambda}
\def\hL{\hat{\Lambda}}
\def\hG{\hat{G}}
\def\hg{\hat{{\cal G}}}
\def\hw{\hat{w}}

\def\O{\Omega}
\def\G{{\cal G }}
\def\S{{\cal S }}
\def\T{{\cal T }}
\def\Z{{\cal Z }}

\def\a{\alpha}
\def\b{\beta}
\def\n{\nu}
\def\m{\mu}
\def\th{\theta}

\def\bbar{\beta}

\def\P{\Phi}
\def\r{\rho}
\def\l{\lambda}
\def\s{\sigma}

\def\e{\epsilon}

\def\t{\tilde}
\def\implies{\Rightarrow}

\begin{document}

\newcommand{\Dirac}{/\!\!\!\!D}
\newcommand{\inv}[1]{{#1}^{-1}} 

\renewcommand{\theequation}{\thesection.\arabic{equation}}
\newcommand{\beq}{\begin{equation}}
\newcommand{\eeq}[1]{\label{#1}\end{equation}}
\newcommand{\ber}{\begin{eqnarray}}
\newcommand{\eer}[1]{\label{#1}\end{eqnarray}}
\begin{titlepage}
\begin{center}
\hfill hep-th/9507064
\vskip .01in \hfill NYU-TH-95/06/01, RI-5-95
\vskip .4in
{\large\bf S-Duality in N=4 Yang-Mills Theories }\footnotemark
\footnotetext{Talk presented by A. Giveon at the {\em Strings '95} conference,
March 13-18, 1995, USC, and the {\em Trieste Conference on S-Duality and
Mirror Symmetry}, June 5-9, 1995.}
\end{center}
\vskip .4in
\begin{center}
{\large Luciano Girardello}$^\clubsuit$,
{\large Amit Giveon}
$^\diamondsuit$,
{\large Massimo Porrati}$^\heartsuit$,
{\large Alberto Zaffaroni}$^\spadesuit$
\vskip .1in
$\clubsuit$ Dipartimento di Fisica, Universit\`a di Milano, via Celoria 16,
20133 Milano, Italy\footnotemark
\footnotetext{e-mail girardello@vaxmi.mi.infn.it}
\vskip .05in
$\diamondsuit$
Racah Institute of Physics, The Hebrew University, Jerusalem
91904, Israel\footnotemark
\footnotetext{e-mail giveon@vms.huji.ac.il}
\vskip .05in
$\heartsuit$ Department of Physics, NYU, 4 Washington Pl.,
New York, NY 10003, USA\footnotemark
\footnotetext{e-mail porrati@mafalda.physics.nyu.edu}
\vskip .05in
$\spadesuit$ Centre de Physique Theorique, Ecole Polytechnique,
F-91128 Palaiseau CEDEX,
France\footnotemark
\footnotetext{e-mail zaffaron@orphee.polytechnique.fr; Laboratoire Propre
du CNRS UPR A.0014}
\end{center}
\vskip .4in
\begin{center} {\bf ABSTRACT} \end{center}
\begin{quotation}
\noindent
Evidence in favor of $SL(2,Z)$ S-duality in $N=4$ supersymmetric Yang-Mills
theories in four dimensions  and with general compact, simple
gauge groups is presented.
\end{quotation}
\vfill
\end{titlepage}
\eject
\def\baselinestretch{1.2}
\baselineskip 16 pt
\noindent
\section{Introduction}
\setcounter{equation}{0}
This talk is based on refs. \cite{GGPZ,GGPZ2}.

Electric-magnetic duality appears already in classical Maxwell's equations
with magnetic monopoles. Here we will present evidence in favor of $SL(2,Z)$
S-duality -- which includes, in particular,
the electric-magnetic duality -- in $N=4$ supersymmetric
Yang-Mills (YM) theories with general, simple gauge groups.

Electric-magnetic duality was conjectured by Montonen-Olive (MO) \cite{MO} for
gauge theories (although the mass spectrum is in general difficult to
compute, due to quantum corrections). A remarkable simplification happens
in $N=4$ supersymmetric gauge theories \cite{WO}: the supersymmetry algebra
implies exact results for masses and charges of ''short multiplets'' --
supersymmetry multiplets containing spin $\leq 1$. The masses are given in
terms of the electric coupling constant, $g_e\equiv g$, and the magnetic one,
$g_m=4\pi/g$, by
\beq
M^2\sim p^2g_e^2+q^2g_m^2,  \qquad p,q\in Z .
\eeq{m2}
Here $M^2$ is invariant under $g_m\leftrightarrow g_e$ together with
$p\leftrightarrow q$. Therefore, $N=4$ is the most likely theory to verify
the MO conjecture \cite{WO,O}.
For general gauge groups, $G$, the electric-magnetic duality transformation
is expected to take the form \cite{GNO}:
\beq
g\to {4\pi\over g}, \qquad G\to\hat{G} ,
\eeq{GGhat}
where $\hat{G}$ is the dual gauge group.

$SL(2,Z)$-duality was recognized in lattice models (with non-zero theta
parameter, $\th\neq 0$) \cite{CR}, and conjectured in string theory
\cite{FIL,SS,Gau}. A version of S-duality is used to compute exact results
in $N=2$ gauge theories \cite{SW}, and a version of electric-magnetic
duality appears also in $N=1$ gauge theories \cite{S}.

Important new evidence for S-duality was found in \cite{Sen,P,VW}.
Sen \cite{Sen} found stable $(p,q)=(2n+1,2)$ states in addition to the
well-known (1,0) electrically charged states, (0,1) monopoles, and the
(1,1) dyons. Recently, it has been shown in ref.~\cite{P}
that {\em all} states with
$p,q$ relatively prime do indeed exist.
A strong-coupling test of S-duality was presented by
Vafa and Witten \cite{VW}. They showed that a topological twisted version of
$N=4$ gauge theories has an S-dual partition function on various manifolds.

Since this is a string theory conference, we should mention that if
S-duality is a fundamental symmetry of string theory, it will explain its
appearance in gauge theories. The outline of the talk is the following.
In section 2, we briefly review the $N=4$ supersymmetric
YM theory. In section 3, we will present
the 't Hooft box with twisted boundary conditions \cite{tH}, generalized to
arbitrary compact $G$, the free energies, and the S-duality conjecture. In
section 4, we will present the result of computing the leading infrared
(IR) divergent term of the free energy, and will discuss its properties. In
section 5, we will present the S-duality transformations of the free
energies. Finally, we will conclude with a few remarks in section 6.

\section{$N=4$ Supersymmetric YM Theories}
\setcounter{equation}{0}
An $N=4$ supersymmetric YM theory is the flat limit ($\a'\to 0$) of
heterotic strings compactified to $D=4$ on a torus\footnote{Recently, in
ref.~\cite{CHL}, it was claimed that there exist
$D=4$, $N=4$ heterotic backgrounds that are {\em not} toroidal
compactifications. Such backgrounds would admit, in particular,
non-simply-laced gauge groups.}. S-duality holds order by order in $\a'$
and, therefore, if it is a symmetry of string theory it is also a
symmetry at the $\a'\to 0$ limit.

$N=4$ supersymmetric YM theory in $D=4$ is completely determined by the
gauge group. The fields in the Lagrangian, $L$, form a supermultiplet:
\ber
(A_{\mu}^a,\,\,\, \l_I^a,\,\,\, \phi_{IJ}^a)& & \nonumber \\
\mu=1,2,3,4, \qquad I,J=1,2,3,4, & & \quad a=1,...,{\rm dim}\, G.
\eer{AlP}
All the fields in (\ref{AlP}) are in the adjoint representation of the
gauge group. The supermultiplet contains a gauge field (spin-1),
$A_{\mu}^a$ ($\mu$ is a space-time vector index and $a$ is a group index of
the adjoint representation),
four Weyl spinors (spin-1/2), $\l_I^a$ ($I$ is the so-called "extension
index,'' in the 4 of $SU(4)$, representing the four supersymmetry charges),
and six scalars (spin-0), $\phi_{IJ}^a$, which obey the condition:
$2\phi_{IJ}^a=\e_{IJKL}(\phi_{KL}^a)^*$.
The Lagrangian takes the form
\ber
L &=& \frac{1}{4\pi}{\rm Re}\, S \Big[{1\over 2}F_{\mu\nu}^aF^{a\,\mu\nu}
+ \bar{\lambda}^{a\, I}\Dirac \lambda_I^a
+ D_{\mu}\phi^{a\, IJ}D^{\mu}\phi^a_{IJ}
\nonumber \\ & &
+ f_{abc}\bar{\lambda}^{a\, I}\phi_{IJ}^b\lambda^{c\, J} +
f_{abc}f_{ade}\phi^b_{IJ}\phi^{c\,JK} \phi^d_{KL}\phi^{e\,LI}\Big]
\nonumber\\ & &
- \frac{i}{8\pi}{\rm Im}\, S F_{\mu\nu}^a\tilde{F}^{a\,\mu\nu}.
\eer{L}
Here $\phi^{a\, IJ}\equiv (\phi^a_{IJ})^*$, $2\tilde{F}^{a\,\mu\nu}=
\e^{\mu\nu\s\r}F^a_{\s\r}$, and $f_{abc}$ are the structure constants of $G$.
{}From $L$ one reads:
\beq
S=\a^{-1}+ia,\qquad \a={g^2\over 4\pi},\qquad a={\theta\over 2\pi},
\eeq{Saa}
where $g$ is the coupling constant and $\th$ is a theta parameter.
The theory is scale invariant: $\b(g)=0$, and the scalar potential has flat
directions when $\langle \P \rangle \in$ Cartan Sub-Algebra (CSA), and is
non-renormalized, even non-perturbatively \cite{seiberg}.

Our aim is to find appropriate gauge invariant quantities which are simple
enough to be calculable, yet non-trivial, i.e., they carry some dynamical
information about the theory, and to test S-duality. One possibility is to
follow 't Hooft strategy where the non-Abelian equivalent of the electric
and magnetic fluxes are defined.

\section{'t Hooft Box with Twisted Boundary Conditions, the Free Energy,
and the S-Duality Conjecture}
\setcounter{equation}{0}

't Hooft  strategy for $SU(N)$ \cite{tH} can be applied to any gauge theory
which contains elementary fields in the adjoint representation and,
in particular, to $N=4$ YM theories. The idea (for $SU(N)$) is to write
Euclidean functional integrals in a box of sides $(a_1,a_2,a_3,a_4)$ with
twisted boundary conditions: $n_{\m\n}\in Z_N$ (the center of $SU(N)$)),
$n_{\n\m}=-n_{\m\n}$. To explain the boundary conditions and generalize to
any compact (simple) $G$ we need some algebra and notations.

The notations are:

\begin{itemize}
\item
$G \equiv$ a compact, simple Lie group.
\item
$\t{G} \equiv$ the universal covering group of $G$.
\item
$G=\t{G}/K$, $K\subseteq C$, $C\equiv Center(\t{G})$.
\item
$\G\equiv$ the Lie algebra of $G$.
\item
$\hG\equiv$ the dual group of $G$.
\item
$\hg\equiv$ the dual Lie algebra, i.e., the Lie algebra of $\hG$.
\item
$\L_R\equiv \L_R(\G)$, the root lattice of $\G$, with normalization
$({\rm long\,\, root})^2=2$.
\item
$\L_W\equiv \L_W(\G)$, the weight lattice of $\G$.
\item
$\hL_R=(\L_W(\G))^{dual}$.
\item
$\hL_W=(\L_R(\G))^{dual}$.
\item
$\hL_{L,R}(\G)=N(\G)\L_{L,R}(\hg)$, where $N(\G)=1$ if $G$ is simply-laced, and
$N(\G)=\sqrt{2}$ if $G$ is non-simply-laced.
\item
The group $G$ has a weight lattice of representations which is a
sub-lattice of $\L_W$: $G=\t{G}/K$ $\implies$ $\L_W(G)=\L_W/K$.
\item
The dual group $\hG$ has a weight lattice dual to the weight lattice of
$G$: $\L_W(\hG)=\L_W(G)^{dual}$.
\item
$\hL(G)_{R,W}=N(\G)\L(\hG)_{R,W}$, where $N(\G)=1$ if $G$ is simply-laced, and
$N(\G)=\sqrt{2}$ if $G$ is non-simply-laced.
\item
For $G$ simply-laced: $\hg=\G$.
\item
For $G$ non-simply-laced: $\G=so(2n+1)\, \Leftrightarrow\, \hg=sp(2n)$.
(The Lie algebrae of $G_2$ and $F_4$ are self-dual.)
\end{itemize}

The center, $C$, of $\t{G}$ is:
\beq
C=\{e^{2\pi i \hw\cdot T} | \hw\in \hL_W/\hL_R\}.
\eeq{Center}
Here $\hw$ is a vector with components $\hw^P$, $P=1,...,r={\rm
rank}\, G$, and $\{T_P\}_{P=1,...,r}$ are the generators in the CSA.
A weight $w=(w_1,...,w_r)$ is the eigenvalue of $(T_1,...,T_r)$
corresponding to one common eigenvector in a single valued representation
of $G$:
\beq
T_PV_w=w_PV_w, \qquad w\in \L_W.
\eeq{Vw}

We now want to evaluate the Euclidean functional integral in a box of sides
$a_{\m}$, with twisted boundary conditions in the center
$\hk_i,\hm_i\in \hL_W/\hL_R\simeq C$, $i,j=1,2,3$ (space indices):
\beq
W[\vhk,\vhm]=\int [dA^a_\mu\, d\lambda^a_I\, d\phi_{IJ}^a] \exp (-\int
d^4x L) .
\eeq{W}
The center elements $\vhk, \vhm$ are defined through the boundary conditions
as follows. The boundary conditions for all bosonic (fermionic) fields are
periodic (anti-periodic) up to a gauge transformation:
\beq
\P(x+a_{\mu}e_{\mu})=(-)^F\O_{\mu}(x)\P(x),
\eeq{PO}
where $e_{\mu}$ is a unit vector in the $\mu$ direction, and repeated
indices are not summed. $\P$ and $\O\P$ denote generically a field of
the supermultiplet (\ref{AlP}) and its gauge transform under $\O$,
respectively; $F$ is the fermion number.
Going from $x$ to $x+a_{\nu}e_{\nu}+a_{\mu}e_{\mu}$, $\mu\neq\nu$,
in two different ways -- either in the $\nu$ direction first and then in the
$\mu$ direction or vice-versa -- implies the consistency conditions:
\ber
\O_{\mu}(x+a_{\nu}e_{\nu})\O_{\nu}(x) &=&
\O_{\nu}(x+a_{\mu}e_{\mu})\O_{\mu}(x)z_{\mu\nu},
\nonumber\\
z_{\mu\nu}&\in& C,
\eer{OOOO}
\beq
z_{\mu\nu}\equiv z_{\hw_{\mu\nu}}=e^{2\pi i\hw_{\mu\nu}\cdot T} , \qquad
\hw_{\mu\nu}\in {\hL_W\over \hL_R}, \qquad  \hw_{\nu\mu}=-\hw_{\mu\nu}.
\eeq{zmn}
The elements $\hm_i$ in $W[\vhk,\vhm]$ are defined by the twists
in the spatial directions:
\beq
\hm_i\equiv {1\over 2}\e_{ijk}\hw_{jk},\qquad i,j,k=1,2,3.
\eeq{m}
$\hm_i$ are interpreted as non-Abelian ``magnetic fluxes'' \cite{tH}.
The elements $\hk_i$ in $W[\vhk,\vhm]$ are defined by the twists in the
time and space directions:
\beq
\hk_i\equiv \hw_{4i}, \qquad i=1,2,3.
\eeq{k}
$\hk_i$ are interpreted as the dual ``electric fluxes.'' The non-Abelian
electric fluxes, $e_i\in \L_W/\L_R$, are linked to $\hk_i$ by the equation:
\beq
e^{-\b F[\ve,\vhm]}={1\over N^3}\sum_{\vhk\in (\hL_W/\hL_R)^3}
e^{2\pi i \ve\cdot \vhk} W[\vhk,\vhm].
\eeq{F}
Here $\b\equiv a_4$ is the inverse temperature, $F[\ve,\vhm]$ is the
free energy of a configuration with electric flux $\ve$ and magnetic flux
$\vhm$:
\beq
\ve=(e_1,e_2,e_3),\,\,\, e_i\in \L_W/\L_R,\qquad  \vhm=(\hm_1,\hm_2,\hm_3),
\,\,\, \hm_i\in \hL_W/\hL_R,
\eeq{vevhm}
and
\beq
N=Order(C\simeq \hL_W/\hL_R).
\eeq{NofC}

The S-duality conjecture is:
\beq
F[\ve,\vhm,1/S,\G]=F[\vhm,-\ve,S,\hg],
\eeq{StoSinv}
\beq
F[\ve,\vhm,S+i,\G]=F[\ve+\vhm,\vhm,S,\G].
\eeq{Sshift}
The transformations $S\to 1/S$ ($\a\to 1/\a$ for $\th=0$, namely,
''strong-weak coupling duality''),  and $S\to S+i$ ($\th\to \th+2\pi$)
generate the S-duality group, isomorphic to $SL(2,Z)$, and acting on $iS$
by:
\beq
iS \rightarrow {a(iS) + b \over c(iS) + d}, \;\;\; a,b,c,d\in Z ,\;
ad-bc=1.
\eeq{i2}
These imply, in particular, that (for simply-laced $G$):
\beq
\Z(1/S,G)=\Z(S,\hG),
\eeq{ZZ}
where
\beq
\Z(S,G)=\sum_{\ve\in (\L_W(G)/\L_R(G))^3,\,\, \vhm\in (\hL_W(G)/\hL_R(G))^3}
e^{-\b F[\ve,\vhm,S]}
\eeq{ZGS}
(recall that for $G=\t{G}/K, K\subseteq C$ , $\L_W(G)=\L_W/K$).

\section{The Free Energy and its Properties}
\setcounter{equation}{0}
In the functional integral representation~(\ref{W}), the integration
over the scalar zero modes, i.e., the VEVs in the Cartan subalgebra, is
divergent. The exact computation of the leading term of such
infrared-divergent $W$,
$w[\vhk,\vhm]$, is presented in detail in ref. \cite{GGPZ2}. Here we shall
only give the result:
\beq
w[\vhk,\vhm]=K[S]\sum_{\hat{w}_{\mu\nu}}
\exp\left[-\pi\sum_{\mu\nu}\left(
\beta V{{\rm Re}\, S \over 2 }{(\hw_{\mu\nu}\cdot \hw_{\mu\nu})
\over a_\mu^2 a_\nu^2} -i{{\rm Im}\, S \over 4}
\epsilon^{\mu\nu\rho\sigma}(\hw_{\mu\nu}\cdot\hw_{\rho\sigma})
\right)\right],
\eeq{m29}
where
\beq
\hw_{ij}=\e_{ijk}(\hl_k+\hm_k), \qquad \hw_{4i}=\hn_i+\hk_i, \qquad
\hk_i,\hm_i\in \hL_W/\hL_R, \,\,\, \hl_i,\hn_i\in \hL_R.
\eeq{klmn}
A convenient choice for the normalization constant $K[S]$ is:
\beq
K[S]=\left( {V(\mbox{Re}\, S)^3\over \b^3}\right)^{r/2},\;\;\;
r=\mbox{rank}\, G.
\eeq{mconst}
After some algebra, one finds from the twisted functional integrals the free
energies:
\beq
\exp\{-\b F[\ve,\vhm,S]\}=c\prod_{i=1}^3\sum_{k_i\in \L_R, \hl_i\in \hL_R}
\exp\Big\{
-\pi\bbar_i(k_i+e_i,\hl_i+\hm_i)M(S)\left(\begin{array}{c}
k_i+e_i \\ \hl_i+\hm_i\end{array}\right)\Big\},
\eeq{FS}
where $c$ is a constant, independent of the fluxes $\ve$ and $\vhm$
and independent of $S$, and
\beq
\bbar_i={\b a_i^2\over V}, \qquad M(S)={1\over {\rm Re}\, S}
\left(\begin{array}{cc} 1 & {\rm Im}\, S \\  {\rm Im}\, S & |S|^2
\end{array}\right)=\a
\left(\begin{array}{cc} 1 & a \\  a & \a^{-2}+a^2 \end{array}\right).
\eeq{MS}

If $\G$ is simply-laced:
\ber
&&\exp\{-\b F[E,M,S,\;{\rm simply-laced}\;\G]\}
=c\prod_{i=1}^3\sum_{K_i^n,L_i^n=-\infty}^{\infty}\exp\Big\{
\nonumber\\
&&
-\pi\bbar_i\Big(K_i^n+(E_iC^{-1})^n,L_i^n+(M_iC^{-1})^n\Big)C_{nm}\otimes M(S)
\left(\begin{array}{c}
K_i^m+(C^{-1}E_i)^m \\ L_i^m+(C^{-1}M_i)^m\end{array}\right)\Big\},
\eer{FEMsl}
where $E_i^n,M_i^n\in Z$, and $C_{nm}$ is the Cartan matrix.
It is remarkable that eq. (\ref{FEMsl}) is formally equal to the classical
piece of a twisted genus-1 string partition function on a toroidal background;
the genus-1 modular parameter is $S$, the target-space background matrix is
$C\otimes I_{3\times 3}$, and the twist is $(E_i,M_i)$.\footnote{This
result could be related to the results reported recently in \cite{other}.}

The free energy obeys factorization, Witten's phenomenon and the 't Hooft
duality.

\begin{itemize}

\item
{\em Factorization}: at $\th=0$:
\beq
F[\ve,\vhm,g,\theta=0]=F[\ve,0]+F[0,\vhm]+c,
\eeq{fac}
where $c$ is independent of the fluxes $\ve$ and $\vhm$.
Factorization is expected to hold in the limit $a_i,\beta\to\infty$, if we
assume that the fluxes occupy only a negligible portion of the total
space \cite{tH}, or if they do not interact, as in the Coulomb phase.
The leading IR-divergent contribution to $F$ is scale invariant:
$F[La_i,L\b]=F[a_i,\b]$ and, therefore, factorization in a large box
implies factorization in any box.

\item
{\em Witten's Phenomenon}: the free energy for $\theta\neq 0$ is derived from
the free energy at $\theta=0$ by the shift:
\beq
e_i\to e_i+{\theta\over 2\pi}\hm_i, \qquad
k_i\to k_i+{\theta\over 2\pi}\hl_i,
\eeq{witten}
Explicitly:
\ber
\exp\{-\b F[\ve,\vhm,S]\}&=&c\prod_{i=1}^3\sum_{k_i\in \L_R, \hl_i\in \hL_R}
\exp\Big\{ \nonumber\\ &&
-\pi\bbar_i(k_i+e_i+{\theta\over 2\pi}(\hl_i+\hm_i),\hl_i+\hm_i)M(g)
\left(\begin{array}{c}
k_i+e_i+{\theta\over 2\pi}(\hl_i+\hm_i) \\ \hl_i+\hm_i\end{array}\right)\Big\},
\nonumber \\
\eer{equa}
where
\beq
M(g)=M(S)|_{\theta=0}=
\left(\begin{array}{cc} {g^2\over 4\pi} & 0 \\  0 & {4\pi\over g^2}
\end{array}\right).
\eeq{Mg}
This is the Witten phenomenon \cite{W}.
Witten's phenomenon also implies that under $\theta\to
\theta+2\pi$, the free energy of electric flux $\ve$ should transform into
the free  energy of electric flux $\ve+\vhm$. For consistency, one can check
that
\beq
e_i+\hm_i\in \L_W,\qquad k_i+\hl_i\in \L_R.
\eeq{cons}

\item
{\em The 't Hooft Duality}:
invariance of $W[\vhk,\vhm]$ under a discrete $O(4)$ rotation:
$1\leftrightarrow 2, 3\leftrightarrow 4$, $m_{1,2}\leftrightarrow k_{1,2}$,
implies
\ber
&&\exp\{-\b F[e_1,e_2,e_3,\hm_1,\hm_2,\hm_3; a_1,a_2,a_3,\b]\}=
{1\over N^2}\sum_{\hk_1,\hk_2\in \hL_W/\hL_R,\, l_1,l_2\in \L_W/\L_R}
\nonumber\\
&&\exp\{2\pi i(\hk_1\cdot e_1+\hk_2\cdot e_2-l_1\cdot\hm_1-l_2\cdot\hm_2)\}
\exp\{-a_3F[l_1,l_2,e_3,\hk_1,\hk_2,\hm_3,a_2,a_1,\b,a_3]\}.
\nonumber \\
\eer{F1234}
This is the 't Hooft duality relation \cite{tH}. Obviously, here there is
nothing to prove since we have computed the functional integrals $w$ and,
therefore, 't Hooft's duality is automatic.

\end{itemize}

\section{S-Duality}
\setcounter{equation}{0}
The S-duality transformations of the free energies are:
\ber
\left(\begin{array}{cc} a & b \\  c & d \end{array}\right)\in SL(2,Z) &:&
\exp\{-\b F[\ve,\vhm,S]\}\to \nonumber\\
& & \exp\{-\b F[\ve,\vhm,(a(iS)+b)/i(c(iS)+d)]\}
\nonumber\\
&=&c\prod_{i=1}^3\sum_{k_i\in \L_R, \hl_i\in \hL_R}
\exp\Big\{
-\pi\bbar_i(k_i+e_i,\hl_i+\hm_i)AM(S)A^t\left(\begin{array}{c}
k_i+e_i \\ \hl_i+\hm_i\end{array}\right)\Big\}\nonumber\\
&=&\exp\{-\b F[d\ve-b\vhm,a\vhm-c\ve,S]\},
\eer{FStrans}
where
\beq
A=\left(\begin{array}{cc} d & -c \\  -b & a \end{array}\right).
\eeq{Aege}
In particular,
\ber
\S=\left(\begin{array}{cc} 0 & -1 \\  1 & 0 \end{array}\right) : S&\to&
{1\over S}, \qquad F[\ve,\vhm,S] \to \nonumber\\
F[\ve,\vhm,1/S]&=&F[\vhm,-\ve,S],
\eer{Str}
i.e., $\S : \G\to \hg$ together with $\ve\to \vm, \vhm\to -\vhe$.
\ber
\T =\left(\begin{array}{cc} 1 & -1 \\  0 & 1 \end{array}\right) : S&\to&
S+i, \qquad F[\ve,\vhm,S] \to \nonumber\\
F[\ve,\vhm,S+i]&=&F[\ve+\vhm,\vhm,S],
\eer{Ttr}
i.e., $\T : \ve\to \ve+\vhm$.

For $\G$ simply-laced, $\hg=\G$ and, therefore,
\beq
F\left[\ve,\vm,{1\over i}{a(iS)+b\over c(iS)+d}\right]=
F[d\ve-b\vm,a\vm-c\ve,S],
\eeq{SofF}
i.e., $F$ is $SL(2,Z)$ covariant.

For $\G$ non-simply-laced, $\G=so(2n+1)\leftrightarrow \hg=sp(2n)$, and
there exist $SL(2,Z)$ transformations that are not allowed (they transform
physical fluxes to unphysical ones). For example,
\beq
\T\S : F[\ve,\vhm,S]\to F[\vhm-\ve,-\ve,S].
\eeq{a3}
But $\vhm-\ve$ is not a vector in $\hL_W^3$ and, therefore, it is an
''illegal'' electric flux in $\hg$.

\section{Summary and Remarks}
\setcounter{equation}{0}

We have defined some gauge invariant quantities in $N=4$ supersymmetric YM
theories based on arbitrary compact, simple groups (the generalization  to
arbitrary compact groups is straightforward): the functional integrals in a
box with twisted boundary conditions, $W[\vhk,\vhm]$, and the corresponding
free energies, $F[\ve,\vhm,S]$. $W[\vhk,\vhm]$ is IR-divergent, and
its leading IR-divergent term is exactly computable.
Therefore, the corresponding leading term of the free energies in all
flux sectors can be derived.

We defined the transformation laws under S-duality of the free energies (the
generalization of the MO conjecture to S-duality in the presence of
non-Abelian fluxes), and we verified that these laws are obeyed by the
quantities we computed. For simply-laced $\G$, $SL(2,Z)$ acts covariantly,
but for {\em non}-simply-laced $\G$, there exist $SL(2,Z)$ transformations
that transform physical fluxes into unphysical ones. Therefore, when $S$ is
promoted to a true dynamical field, $SL(2,Z)$ is not a true symmetry (but
only a sub-group) if $G$ is non-simply-laced. Such gauge groups can never
be obtained from $N=4$ toroidal compactifications of the heterotic string
and, therefore, in the moduli space of $N=4$ toroidal compactifications,
$SL(2,Z)$ S-duality is expected to be a symmetry\footnote{If different $N=4$
heterotic string backgounds, which are not
equivalent to toroidal compactifications, and admit non-simply-laced
gauge groups exist, as claimed in ref.~\cite{CHL},
only a subgroup of $SL(2,Z)$, described in \cite{GGPZ2},
is expected to be a symmetry in the moduli space of such backgrounds.}.

Now, it is time to discuss the partition function and electric-magnetic
duality. For a gauge {\em group}, $G$, not all flux sectors are permitted,
but only:
\beq
e_i\in {\L_W(G)\over \L_R(G)}, \qquad
\hat{m}_i\in {\hL_W(G)\over \hL_R(G)}.
\eeq{minLW}
Recall that $\L_W(G)=\L_W/K$ for $G=\t{G}/K$, $K\subseteq C$ (see section
3 for the other notations). As mentioned before, the partition function,
$\Z(S,G)$, is given by summing over all allowed flux sectors:
\beq
\Z(S,G)=\sum_{\ve\in (\L_W(G)/\L_R(G))^3,\,\, \vhm\in (\hL_W(G)/\hL_R(G))^3}
e^{-\b F[\ve,\vhm,S]}.
\eeq{ZGS2}
The electric-magnetic duality is
\ber
\S &:& \,\, \Z(S,G)\to \Z(1/S,G)=\Z(S,\hG), \qquad\,\,\,\,\, {\rm if}\,\,\,
G \,\, {\rm simply-laced} \nonumber\\
\S &:& \,\, \Z(S,G)\to \Z(1/S,G)=\Z(S/2,\hG), \qquad {\rm if}\,\, G \,\, {\rm
non-simply-laced}.
\nonumber \\
\eer{SofZ}

The partition function $\Z$ is invariant under the subgroup of $SL(2,Z)$
generated by $\{\T^n,\S\T^{\hn}\S \}$, where $n\in Z$ such that
$e_i+n\hm_i\in \L_W(G)$ for any $e_i\in \L_W(G)/\L_R(G), \hm_i\in
\hL_W(G)/\hL_R(G)$, and $\hn\in Z$ such that $\hm_i+\hn e_i\in \hL_W(G)$ for
any $e_i\in \L_W(G)/\L_R(G), \hm_i\in \hL_W(G)/\hL_R(G)$.

For example, if $\G=su(2)$:
\beq
\Z(SU(2))=\sum_{e_i=0,1/\sqrt{2}} e^{-\b F[\ve,0]}={1\over 4} W[0,0].
\eeq{ZSU2}
It is invariant under the subgroup $\Gamma_0(2)$ generated by
$\{\T,\S\T^2\S\}$. The partition function of the dual group
$\hat{SU}(2)=SO(3)\simeq SU(2)/Z_2$ is
\beq
\Z(SO(3))=\sum_{m_i=0,1/\sqrt{2}} e^{-\b F[0,\vm]} =
{1\over 8} \sum_{k_i,m_i=0,1/\sqrt{2}} W[\vk,\vm].
\eeq{ZSO3}
Under electric-magnetic duality, indeed,
\beq
\S : \Z(SU(2)) \leftrightarrow \Z(SO(3)).
\eeq{SSU2}

Back to the general case, we should remark that in the Hamiltonian
formalism we can evaluate $F[\ve,\vhm]$ at $\th=\vhm=0$, $g\ll 1$. Then, by
imposing factorization, the 't Hooft duality and the Witten phenomenon, one
can {\em derive} the result for $\th,\vhm\neq 0$ from the $\th=\vhm=0$ one.
The constant scalar fields and gauge
fields modes in the CSA, $(\phi_{IJ},c_i)$, live on the orbifold~\cite{IM}:
\beq
(\phi_{IJ},c_i)\in {R^{6r}\times \prod_{i=1}^3 T_i^r \over {\rm
Weyl-Group}}, \qquad T_i^r\equiv {R^r\over 2\pi\hL_R/a_i}, \qquad r={\rm
rank}\, G .
\eeq{Zaf2'}
More generally,
Imbimbo and Mukhi showed how to take into account the scalar divergence in
the Hamiltonian approach.

To conclude, we remark that further highly non-trivial tests of S-duality
in $N=4$ supersymmetric YM theories could be done by computing the
subleading terms in the IR divergence expansion. Moreover, it would be
interesting to see if similar tests can be applied also to $N<4$
supersymmetric gauge theories.

\vskip .5in \noindent
{\bf Acknowledgements} \vskip .2in \noindent
L.G. is supported in part by the
Ministero dell' Universit\`a e della Ricerca Scientifica e Tecnologica,
by INFN and by ECC, contracts SCI*-CT92-0789 and CHRX-CT92-0035.
A.G. is supported in part by BSF - American-Israel Bi-National
Science Foundation, by the BRF - the Basic Research
Foundation, and by an Alon fellowship. M.P. is supported in part by NSF
under grant no. PHY-9318171. A.Z. is supported in part by ECC, Project
ERBCHGCTT93073, SCI-CT93-0340, CHRX-CT93-0340.

\end{document}